| | |
|---|---|
| **Title** | **From Repeatability to Self-Organization of Guided Streamers Propagating in a Jet of Cold Plasma** |
| **Authors** | H. Decauchy[1], T. Dufour[1] |
| **Affiliations** | [1]LPP, Sorbonne Université Univ. Paris 6, CNRS, Ecole Polytech., Univ. Paris-Sud, Observatoire de Paris, Université Paris-Saclay,10 PSL Research University, 4 Place Jussieu, 75252 Paris, France. |
| **Correspondence** | E-mail: thierry.dufour@sorbonne-universite.fr |
| **Ref.** | Plasma, Vol. 6, pp. 250-276 (2023) |
| **DOI** | https://doi.org/10.3390/plasma6020019 |
| **Abstract** | In this work, a jet of cold plasma is generated in a device supplied in helium and powered with a high-voltage nanopulse power supply, hence generating guided streamers. We focus on the interaction between these guided streamers and two targets placed in a series: a metal mesh target (MM) at floating potential followed by a metal plate target (MP) grounded by a 1500 Ω resistor. We demonstrate that such an experimental setup allows to shift from a physics of streamer repeatability to a physics of streamer self-organization, i.e., from the repetition of guided streamers that exhibit fixed spatiotemporal constants to the emergence of self-organized guided streamers, each of which is generated on the rising edge of a high-voltage pulse. Up to five positive guided streamers can be self-organized one after the other, all distinct in space and time. While self-organization occurs in the capillary and up to the MM target, we also demonstrate the existence of transient emissive phenomena in the inter-target region, especially a filamentary discharge whose generation is directly correlated with complexity order Ω. The mechanisms of the self-organized guided streamers are deciphered by correlating their optical and electrical properties measured by fast ICCD camera and current-voltage probes, respectively. For the sake of clarity, special attention is paid to the case where three self-organized guided streamers ($\alpha$, $\beta$ and $\gamma$) propagate at $v_\alpha$ = 75.7 km·s⁻¹, $v_\beta$ = 66.5 km·s⁻¹ and $v_\gamma$ = 58.2 km·s⁻¹), before being accelerated in the vicinity of the MM target. |
| **Keywords** | self-organization; guided streamers; emergence; transient emissive phenomena; filamentary discharge; complexity order; leaders; plasma jet; physics of repeatability |

# 1. Introduction

## 1.1. Atmospheric Pressure Plasma Jets

Atmospheric pressure plasma jet (APPJ) devices emerged at the beginning of the 1960s and have been used to generate high temperature plasmas, several thousand Kelvin, for applications such as diamond synthesis, metal etching, thermal spraying and ceramic coatings deposit [1–3]. In parallel, much fundamental research has been devoted to the characterization of their thermal properties as well as the conditions governing their local thermodynamic equilibrium [4,5]. Later, in the 1990s, APPJs operating out of thermodynamic equilibrium appeared, characterized by gas temperature comprised between 25 and 200 °C, charged particle densities of approximately $10^{11}$–$10^{12}$ cm⁻³ and reactive species concentrations as high as 10 to 100 ppm [6]. Since APPJ can be scaled up to treat large areas, they can address many applications that were previously limited to low pressure processes, e.g., deposition of silicon oxide films, polymer etching, surface hydrophilization, etc. [7]

Today, these plasma jet devices can be used in ambient air, paving the way for tremendous applications in the life sciences. This is the case in medicine where APPJs can be used in dentistry for the sterilization of tooth root canal but also in oncology to induce antitumor effects and trigger immune responses [8–10]. In agriculture, APPJs can be utilized to achieve small-scale treatments such as the activation of small water samples to promote plants growth and could be envisioned as a tool for precision agriculture [11,12]. In all these studies, the APPJ devices propagate the cold plasma from the high-voltage electrode to the capillary outlet as ionization waves called streamers.

## 1.2. Streamers

A streamer can be defined as an ionization wave that propagates longitudinally to carry electrical charges and radiative species over long distances, from a few mm to several cm away from the interelectrode region [13]. The most emissive region of the streamer is its head whose ionization front is characterized by a very intense electric field. There, electrons can reach typical energies of the order of 10 eV or more [14], triggering chemical reactions through the production of various active species:

- Non-reactive species that can be excited to upper energetic levels (e.g., nitrogen molecules in the second positive system, Helium metastable species, …)
- Chemically reactive species, whether long-lived (e.g., $O_3$) or very short-lived (e.g., radicals with one or more unpaired electrons on their outer layer such as OH, NO, O, O*).

Depending on the type of plasma sources, streamers can be generated with stable or unstable spatiotemporal properties:

- In nano- or micro-pulsed plasma discharges, streamers can be repeated periodically in space and time: they follow each other at regular time intervals and are formed at the same spatial coordinates. Generally, these so-called guided streamers are produced by high-voltage pulses with widths of a few µs and periods of a few tens of ms [15,16].

- In dielectric barrier devices supplied by AC power supply or in point-plate configurations supplied by DC voltage, streamers are generated with random spatial and temporal distributions. Typically, these streamers follow each other at irregular time intervals and without a predictable spatial arrangement. In the latter case, a distinction can be made between threadlike streamers (which remain single streamers while gradually decreasing in diameter and velocity without branching) and branched streamers (which split into thinner streamers to form a tree-like structure). These streamers generally branch into two new channels that move away from each other since neighboring streamers generally have the same polarity (repulsion forces) [17].

Whether the streamers are spatiotemporally guided, randomly distributed or branched, characterizing their dynamics and interactions is of major interest to decipher fundamental processes such as self-organization.

## 1.3. Self-Organization

Self-organization is a process by which the organization (constraint, redundancy) of a system spontaneously increases, i.e., without this increase being controlled by the environment or an encompassing system or an otherwise external system [18]. It can also be defined as a dynamical and adaptive process where systems acquire and maintain structure themselves, without external control [19]. These definitions stem from historical principles, such as "order from noise" formulated by Heinz von Foerster (1960) [20], "complexity from noise" stated by Henri Atlan (1977) [21], as well as "order through fluctuations" and "order from chaos" postulated by Ilya Prigogine [22,23]. They are illustrated by famous examples such as flocks of birds self-organized into V-formations for traveling long distance [24], bridges built by army ants with their own bodies to accelerate their foraging excursions [25], sandpiles formed by pouring grains on a flat surface that collapse by avalanche then grow and collapse again, and so on [26]. In that later case, the resulting "self-organized criticality" applies whether the grains of sand are poured simultaneously or successively (grain after grain), whether they are poured randomly or regularly over time. In physics, many examples illustrate self-organization processes, including arrangement of ferromagnetic objects and films [27,28], quantum particles at zero temperature in optical lattice [29], systems near the rigidity percolation threshold [30], etc. In the specific research field of cold plasma physics, self-organized patterns have also been highlighted in cathode boundary layer micro-discharges [31], DC pin electrodes interacting with liquids [32], high power impulse magnetron sputtering [33].

In this article, a plasma jet device is supplied with helium and polarized to high-voltage pulses to generate guided streamers. The interaction of these streamers with two targets placed in parallel leads to a self-organization of the guided streamers, which will be highlighted and analyzed.

# 2. Materials and Methods

## 2.1. CHEREL Device and Targets

In this work, guided streamers are generated by a plasma device corresponding to a Capillary with inner Hollow Electrode and outer Ring ELectrode (CHEREL). The capillary, in quartz material, is 150 mm in length and has inner and outer diameters of 2 mm and 4 mm, respectively. The inner hollow electrode (copper material, $d_{in}$ = 1 mm, $d_{out}$ = 2 mm, length = 50 mm) is coaxially centered in the capillary and connected to the high voltage while the outer ring (aluminum material, $d_{in}$ = 2 mm, $d_{out}$ = 4 mm, length = 10 mm) is grounded, as indicated in **Figure 1a**.

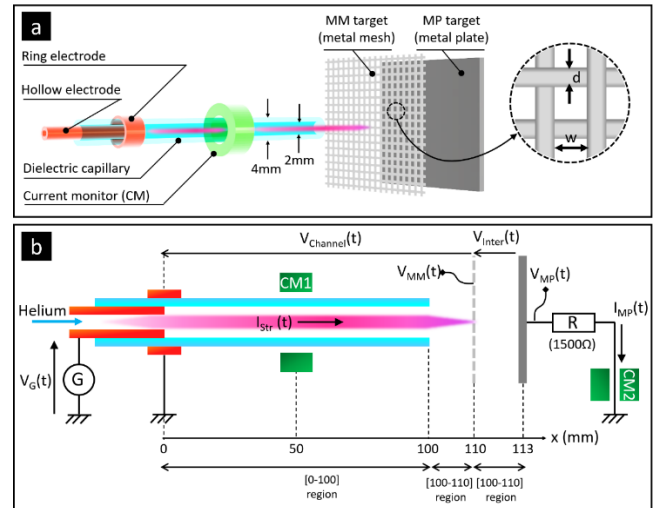

*Figure 1. (a) CHEREL-MM-MP configuration. The CHEREL (Capillary with inner Hollow Electrode and outer Ring ELectrode) generates a jet of cold plasma that interacts with MM (metal mesh target) and MP (metal plate target) placed in series, (b) Equivalent electrical schematics of (a) including voltage generator, electrical probes and resistor for electrical characterizations.*

The CHEREL is supplied with helium at 1000 sccm and supplied with positive pulses of high voltage generated by a pulse generator (RLC electronic Company, NanoGen1 model) coupled with a DC high-voltage power supply (Spellman company, SLM 10 kV 1200 W model). In all the experiments, the high-voltage magnitude is comprised between 4.5 and 5.5 kV, the duty cycle is comprised between 10 and 15 % and the repetition frequency is 5 kHz.

The **Figure 2a** represents an ideal voltage pulse where rise time ($\tau_{rise}$) and fall time ($\tau_{fall}$) are equal to 0. Conversely, **Figure 2b** corresponds to a real pulse supplied by the generator (A = 5000 V, $f_{rep}$ = 5 kHz and $D_{Cycle}$ = 10%) and where the amplitude exceeds the set value by at least 400 V. **Figure 2c,d** show enlargements of the rising and falling edges, respectively, so as to highlight the characteristic times, $\tau_{rise}$ and $\tau_{fall}$, both estimated at 20 ns. Compared to the setpoint of 5 kV, the measured amplitude is 2.47 kV higher on $\tau_{rise}$ and 2.33 kV higher on $\tau_{fall}$.

The CHEREL is completed by two targets, as sketched in **Figure 1a**:

- A metal mesh target (MM) at floating potential placed 10 mm from the dielectric capillary. Surface area of the MM target is 8 cm × 8 cm and its mesh has a nominal aperture (w) of 300 μm, a wire diameter (d) of 220 μm and an estimated void percentage of $100 . \left( \frac{w}{w+d} \right)^2 = 33\%$.
- A metal plate target (MP) connected to a 1.5 kΩ resistor that is grounded. Surface area of the metal plate is 8 cm × 8 cm. It is placed 3 mm from the mesh, as shown in **Figure 1b**.

Three regions of cold plasma propagation can thus be distinguished: between the polarized electrode and the capillary outlet, between the capillary outlet and the MM target and the so-called "inter-target region" comprised between the MM and MP targets. Finally, it should be noted that these two targets can be considered as the electrodes of a same capacitor whose dielectric medium is the ambient air. The MM/air/MP region behaves, therefore, like a series RC circuit of time constant $\tau_{RC} = R.C.$

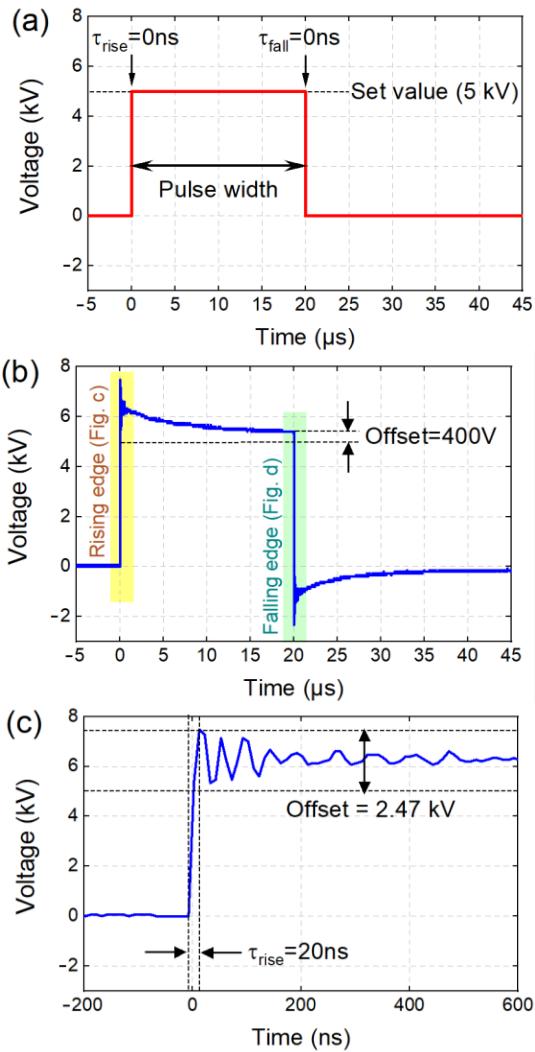

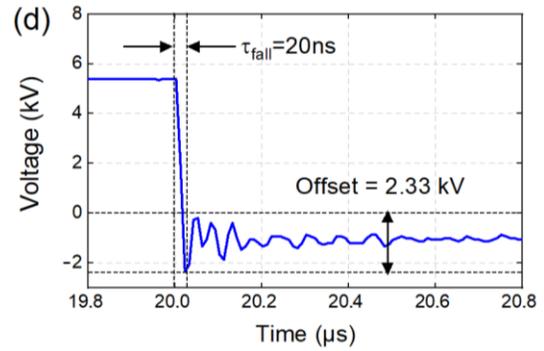

*Figure 2. (a) Ideal positive square pulse, (b) real positive square pulse obtained with the HV power supply, (c) enlarged view of the pulse rising edge to evidence $\tau_{rise}$, overshoot and ringing, (d) enlarged view of the pulse falling edge to evidence $\tau_{fall}$ and backswing.*

## 2.2. Diagnostics

### 2.2.1. Electrical Characterizations

Electrical characterization of the guided streamers is achieved using voltage and current probes connected to an oscilloscope (Model Wavesurfer 3054 from Teledyne Lecroy company, Chestnut Ridge, NY, United States). As sketched in **Figure 1b**:
- A current monitor (Model 2877, from Pearson Electronics, Beccles, United Kingdom) is coaxially centered around the dielectric capillary at x = 50 mm to measure the current of the guided streamers ($I_{Str}$).
- A high-voltage probe (Model P6015A 1000:1, from Tektronix, Beaverton, OR, United States) is connected to the MM target to measure its voltage ($V_{MM}$).
- A current monitor (Pearson, 2877) and a high-voltage probe (Model PPE 20 kV 1000:1, from Teledyne LeCroy) are connected to the MP target to measure its current ($I_{MP}$) and voltage ($V_{MP}$), respectively.

In addition, a LCRmeter (HM-8118 model from Rohde & Schwarz, Munich, Germany) is utilized to measure the capacitance of the inter-target region.

### 2.2.2. Fast ICCD Imaging and Optical Emission Spectroscopy

Since guided streamers are transient and low emissive phenomena, the observation of a single streamer would require a streak camera with a temporal resolution close to (or better than) 800 fs [34]. Since this equipment is not in our facilities, an alternative is to study a train of guided streamers, i.e., collecting at regular time intervals a large number of guided streamers, here 25 000, and summing their emissions on a single image. The radiative emission of the plasma jet is therefore collected by an intensified charge-coupled device (ICCD) camera from Andor company (model Istar DH340T). It has a 2048 × 512 imaging array of 13.5 μm × 13.5 μm pixels and an optical gate width lower than 2 ns. The Solis software enables such operations combining the "kinetic series" acquisition mode and the "DDG" gate mode. More information, especially concerning the calibration procedure, is detailed in [15].

Optical emission spectroscopy is achieved to identify the radiative species from plasma. The spectrometer (SR-750-B1-R model from Andor company, Belfast, United Kingdom) operates in the Czerny Turner configuration, with a 750 mm focal length while diffraction is achieved with a 1200 grooves·mm⁻¹ grating in the visible range. It is equipped with the same camera as the one used to achieve fast ICCD imaging.

# 3. Results and Discussion

## 3.1. Experimental Conditions Driving to Self-Organization of Guided Streamers within a Plasma Jet

### 3.1.1. Emergence of Self-Organized Guided Streamers

To unambiguously differentiate streamer repeatability from streamer self-organization, three configurations are realized as shown in **Figure 3**: (i) configuration where the CHEREL interacts only with the MP target connected to the 1.5 kΩ resistor, (ii) configuration where the CHEREL interacts only with the MM target at floating potential and (iii) configuration where the CHEREL interacts with the two targets placed one after the other. Supplied with helium and powered by high-voltage pulses, the CHEREL generates a cold plasma jet that extends several cm away from the inter-electrode gap. When photographed with a conventional camera for an exposure time of 1 s and a numerical aperture of 4.5, a continuous emissive jet is observed, whether interacting with MP (**Figure 3a**), MM (**Figure 3b**) or both targets (**Figure 3c**). This optical artifact disappears using a fast ICCD camera, which allows to visualize the unitary elements of the plasma jet: the "guided streamers" that follow one another at a frequency of 5 kHz. In the CHEREL-MP and CHEREL-MM configurations, the trains of guided streamers are superimposed at the same spatial coordinates, thus revealing only a single pattern, commonly (but abusively) called bullet. On the other hand, in the CHEREL-MM-MP configuration, this superposition of guided streamers gives rise to three spatially distinct and thus self-organized patterns.

This same phenomenon can be demonstrated by electrical characterization, as proposed in **Figure 4**, where each current peak does represent a single guided streamer and not a train of guided streamers. In the CHEREL-MP configuration, the current peaks repeat periodically with the same amplitude, whether measured in the capillary (**Figure 4d**) or in the target (**Figure 4j**). Similarly, the voltage measured in the target corresponds to a succession of identical pulses, 20 μs wide, repeated at a frequency of 5 kHz (**Figure 4g**). In the CHEREL-MM configuration, $V_{MM}(t)$ presents a similar profile: pulses of amplitude varying between 1.9 and 3.0 kV that repeat periodically (**Figure 4b**). Unsurprisingly, **Figure 4e** shows that the positive current peaks (associated with the rising edges of the high-voltage pulses) are repeated identically; so are the negative current peaks (associated with the falling edges). In contrast to these two situations where the plasma jet interacts with a single target, the voltages associated with the CHEREL-MM-MP configuration can present three different amplitude levels at each period ($T_G$ = 1/5kHz = 200 ms) whether for $V_{MM}(t)$ or $V_{MP}(t)$, in **Figure 4c,i**, respectively.

If the repeatability in time and amplitude is observed for the CHEREL-MP and CHEREL-MM configurations, only the repeatability in time remains in the third configuration. Indeed, the CHEREL-MM-MP configuration sees the amplitude of its voltage and current signals structured on three levels instead of one. This greater complexity of the signal is the signature of a self-organization phenomenon occurring within the plasma jet. Interestingly, it is not only limited to three patterns, as detailed in the next section.

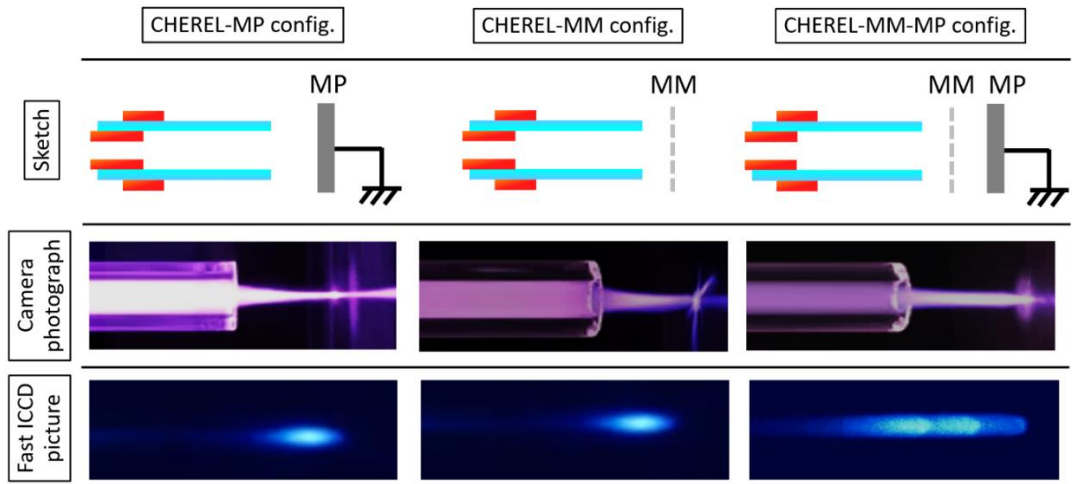

*Figure 3. Pictures obtained by digital photography and by fast ICCD imaging, considering the CHEREL-MP configuration (train of repeated guided streamers), the CHEREL-MM configuration (train of repeated guided streamers) and the CHEREL-MM-MP configuration (train of self-organized guided streamers).*

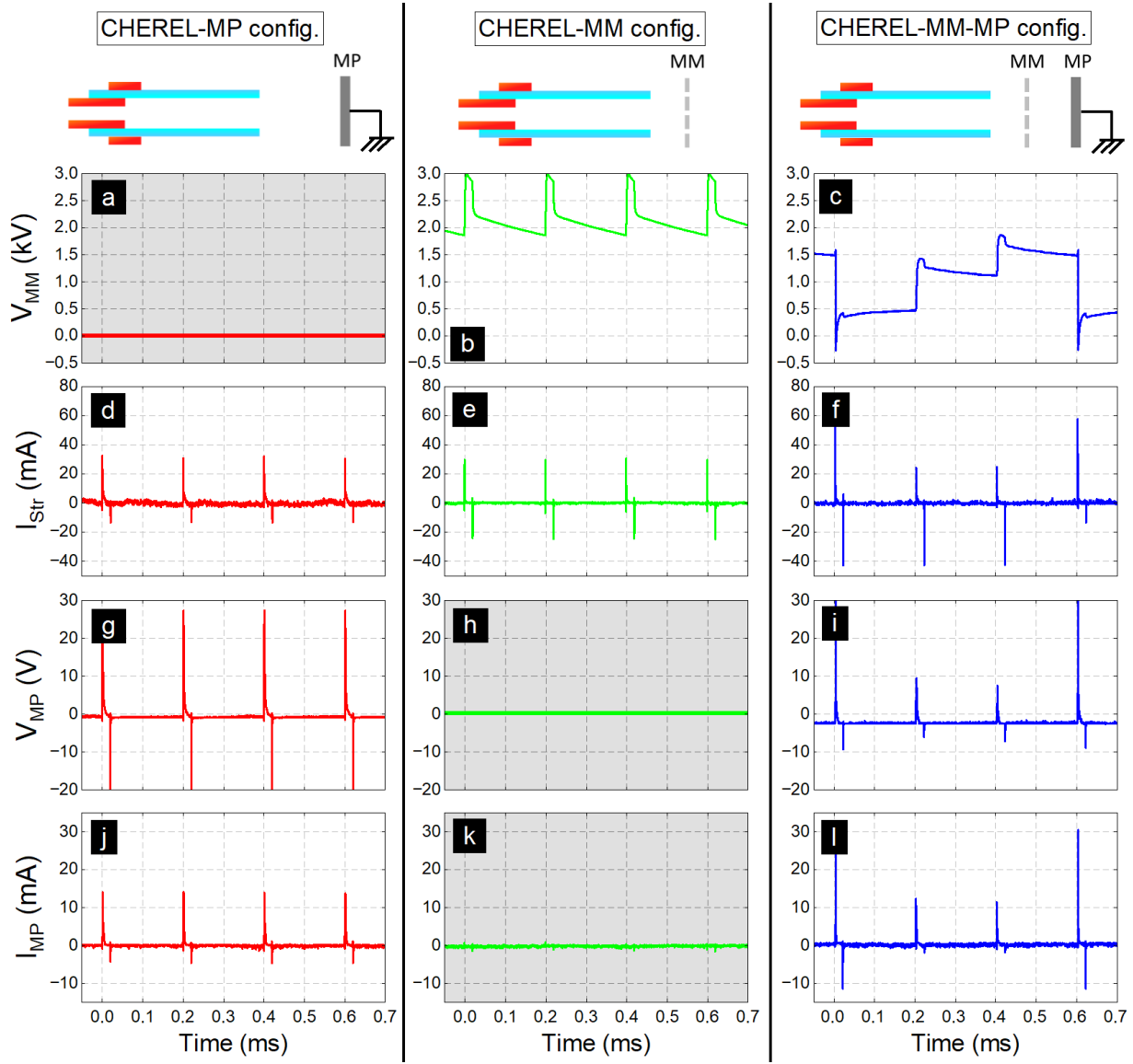

*Figure 4. Electrical characterizations to distinguish repeatability and self-organization of guided streamers within a plasma jet considering three configurations: (a,d,g,j) the CHEREL-MP configuration characterized by repeatability, (b,e,h,k) the CHEREL-MM configuration characterized by repeatability, (c,f,i,l) the CHEREL-MM-MP configuration characterized by self-organization.*

### 3.1.2. Complexity Order

Complexity order ($\Omega$) is the number of unitary patterns capable of self-organizing within a same system. This parameter distinguishes the absence of self-organization ($\Omega = 1$) from self-organization itself ($\Omega > 1$), which can gain complexity ($\Omega \gg 1$).

In the CHEREL-MM-MP configuration, **Figures 3** and **4** show that three consecutive guided streamers can self-organize (i.e., $\Omega = 3$). Further complexity orders can be obtained by only slightly changing the values of the electrical parameters (amplitude, frequency, duty cycle) as reported in **Table 1** and as illustrated in **Figure 5** with $\Omega = 1$, 2, 3 and 5. **Figure 5a** corresponds to the situation classically encountered in the literature: a simple repeatability phenomenon without self-organization ($\Omega = 1$). This is a train of 25 000 guided streamers that all overlap on the same ICCD image. The corresponding $V_{MM}(t)$ voltage profile is structured over two states: the low state at 0 V (reference) and the high state ($\alpha$ state) near 2 kV. The correlation between the fast ICCD imaging and the electrical characterization makes it possible to state that these guided streamers are generated on the rising edge of each applied voltage pulse [15]. In **Figure 5b**, the guided streamers are self-organized into two distinct patterns ($\Omega = 2$) and the related electrical characterization associates these patterns to the $\alpha$ and $\beta$ states at about 0.7 kV and 1.7 kV, respectively. A complexity order of three is highlighted in **Figure 5c** where the voltage measured on the MM target makes it possible to distinguish a signal with three distinct and non-zero states: α, β and γ at about 0.5 kV, 1.1 kV and 1.4 kV, respectively. Fast ICCD imaging represents these three states as spatially dissociated patterns of guided streamers, with the α streamer appearing slightly less bright than the β and γ streamers. Finally, **Figure 5d** shows a self-organization phenomenon of order five, correlating on the one hand a fast ICCD picture where the corresponding patterns (α, β, γ, δ and ε) can be discerned, and on the other hand a voltage signal with five non-zero electrical states, respectively, at 0.5 kV, 1.0 kV, 1.3 kV, 1.6 kV and 1.7 kV.

While guided streamers can self-organize into Ω patterns, we still ignore whether these are induced by positive guided streamers

(triggered on the rising edges of the pulse, see **Figure 2**) and/or negative guided streamers (triggered on the falling edges of the same pulse). In the next section, we will therefore distinguish between positive guided streamers in the states α+, β+, γ+, etc. and negative guided streamers in the states α−, β−, γ−, etc. The terms of α phase, β phase, γ phase, etc. will be used to designate the time intervals during which the states {α+, α−}, {β+, β−}, {γ+, γ−}, etc. appear. In the remainder of this paper and for the sake of clarity, the results and discussion will focus on the order Ω = 3 (although all our explanations can obviously be extended to other values of Ω).

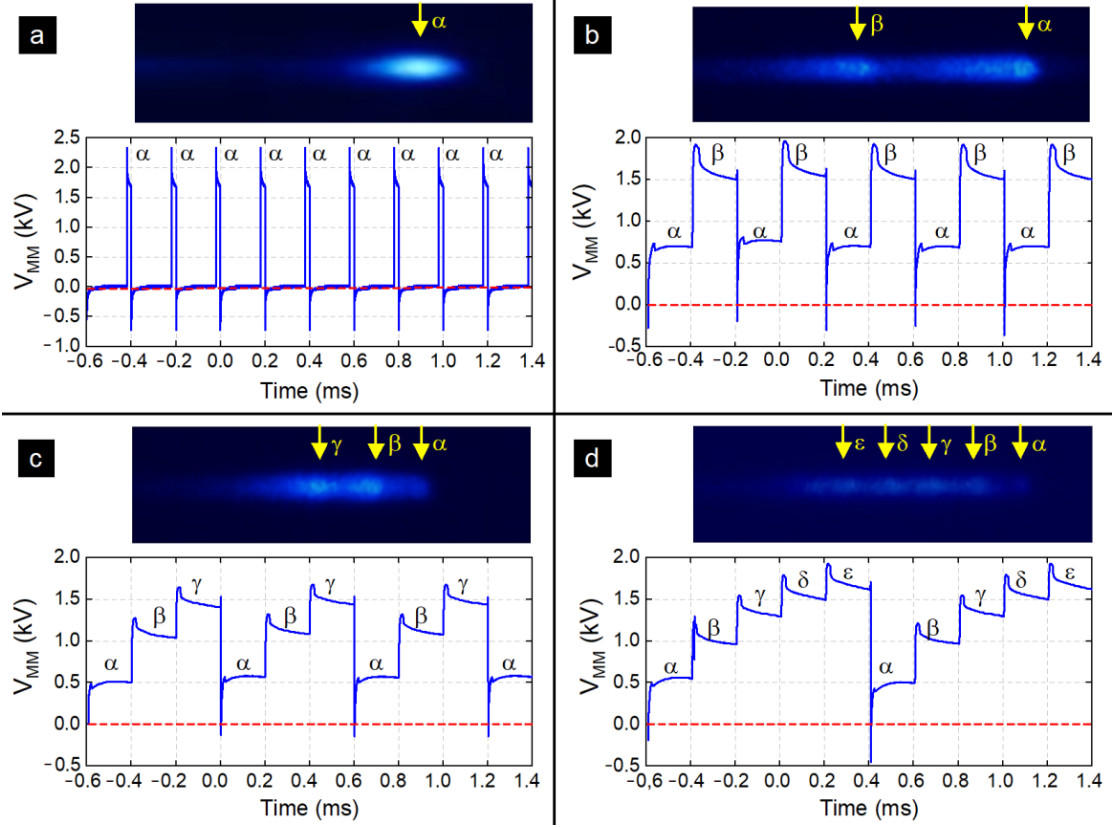

*Figure 5. Spatial profiles of self-organized guided streamers measured by fast ICCD imaging and time profiles of the voltage measured on the MM target for (a) Ω = 1, i.e., no self-organization, (b) Ω = 2, (c) Ω = 3 and (d) Ω = 5. The symbols α, β, γ, δ and ε represent the self-organizing states in order of appearance.*

| Ω | Amplitude (kV) | Frequency (kHz) | Duty Cycle (%) |
|---|---|---|---|
| 1 | 6.0 | | 15 |
| 2 | 5.5 | | |
| 3 | 5.0 | 5 | |
| 4 | 4.7 | | 10 |
| 5 | 4.5 | | |
| 6 | 4.3 | | |

*Table 1. Complexity orders (Ω) and their associated electrical parameters in the case of CHEREL interacting with MM and MP targets, as sketched in Figure 1.*

## 3.2. Optical Characterization (Ω = 3)

The propagation properties of the self-organized guided streamers are studied by fast ICCD imaging for Ω = 3, in the CHEREL-MM-MP configuration. The study focuses on both the positive and negative streamers but also on the transient emissive phenomena (TEP) induced by the α, β and γ streamers.

### 3.2.1. Positive Guided Streamers

Positive guided streamers, generated on the rising edges of voltage pulses, can be subject to propagation and counter-propagation phenomena when they interact with a target [15,35]. A train of 25 000 positive guided streamers, overlaid on the same ICCD picture, is monitored in **Figure 6**, with enlarged views at the capillary's outlet and in the inter-target region. We will explain how such study can provide information on spontaneous emergence of self-organization, streamer propagation velocities and duration between consecutive streamers. Additionally, we will see that a space–time emission diagram can be plotted (**Figure 7**) by transversally summing the pixels of each fast ICCD picture.

Spontaneous Emergence of Self-Organization

The train of guided streamers in **Figure 6** is initiated at t = 0 ns from the HV electrode, then propagates in the capillary, passes through the current probe at 340 ns and exits at 600 ns. In the immediate vicinity of this probe, all the guided streamers spatially overlaid at regular time intervals ($T_G = 1/f_{rep}$ = 200 ms), forming a single pattern. Between 50 and 70 mm (i.e., 600 ns < t < 1000 ns), this pattern starts to lengthen but it is only from x = 70 mm, that it splits into three; this separation becomes particularly pronounced between 1060 and 1300 ns. The self-organization of these three patterns (i.e., trains of positive guided streamers) emerges spontaneously (i.e., without external constraint) and progressively (i.e., the spatial and temporal separation is not instantaneous). Finally, the α, β and γ streamers reach the MM target at 1370 ns, 1440 ns and 1540 ns, respectively.

Propagation Velocities

The propagation velocities of the guided streamers can be deduced from **Figure 7** considering the slopes of the black dashed lines. The corresponding values are reported in **Table 2**. It appears that:

- In the capillary, the α streamers are the fastest (75.7 km·s⁻¹), followed by the β (66.5 km·s⁻¹) and γ (58.2 km·s⁻¹) streamers.
- At the capillary's outlet, the three self-organized guided streamers gain in velocity while keeping the same inequality ratios ($v_α > v_β > v_γ$).
- A more detailed explanation on the difference of velocities between α, β and γ is proposed in Section 3.3.2.

| Streamers | In the Capillary (75 < x < 100 mm) | Region between Capillary and MM Target |
|---|---|---|
| α | 75.7 km·s⁻¹ | 159.6 km·s⁻¹ |
| β | 66.5 km·s⁻¹ | 140.3 km·s⁻¹ |
| Γ | 58.2 km·s⁻¹ | 124.2 km·s⁻¹ |

**Table 2.** Velocity of the α, β and γ streamers measured by fast ICCD imaging either in the capillary or in the region separating the capillary from the MM target (A = 5 kV, $f_{rep}$ = 5 kHz, $D_{Cycle}$ = 10 %).

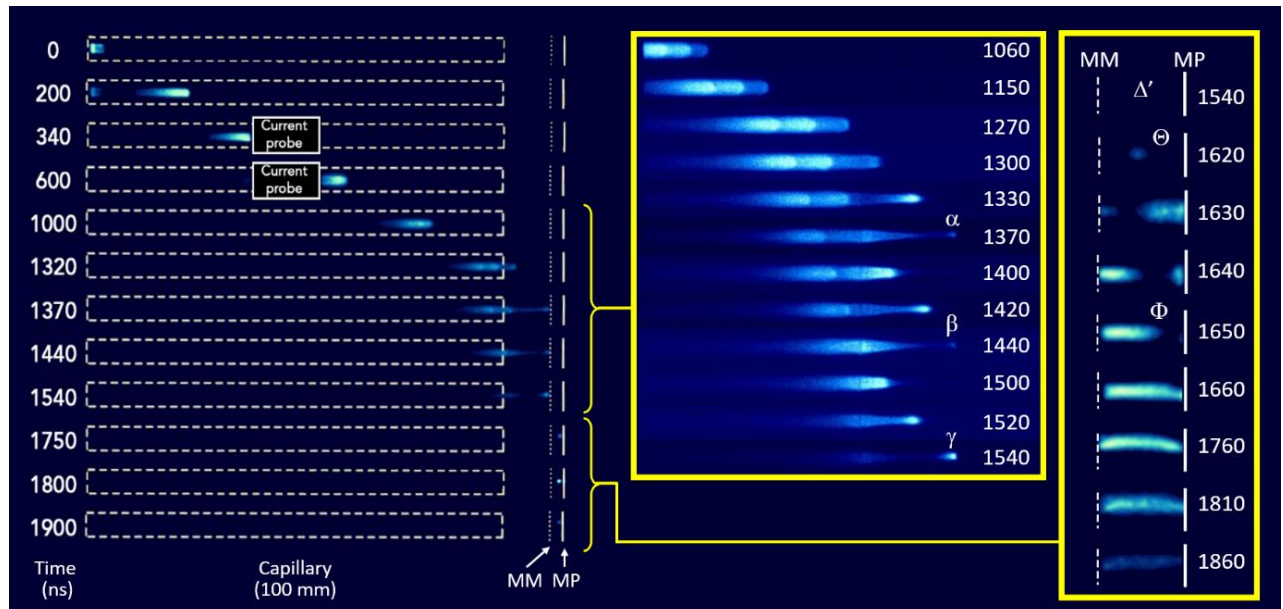

*Figure 6. Fast ICCD pictures of positive guided streamers (Ω = 3) propagating in the CHEREL's capillary, then interacting with ambient air, metal mesh (MM) target at floating potential and metal plate (MP) target connected to R=1.5 kΩ that is grounded (A = 5 kV, $f_{rep}$ = 5 kHz, $D_{Cycle}$ = 10%).*

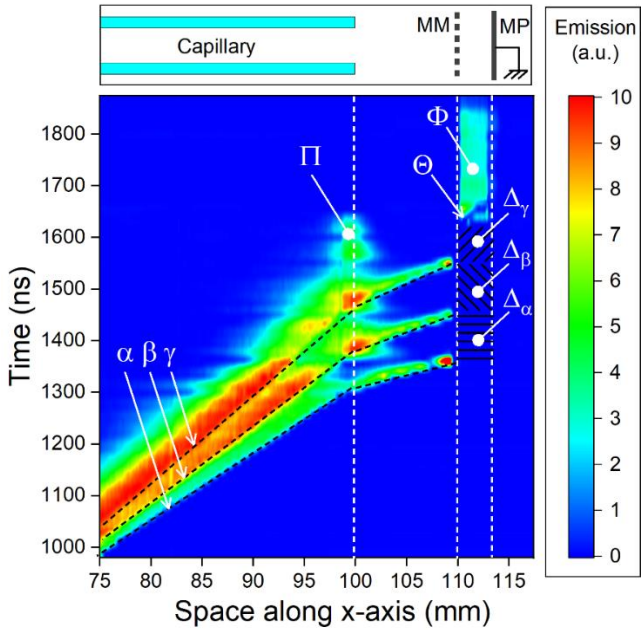

*Figure 7. Space–time emission diagram obtained from fast ICCD imaging to highlight the self-organized guided streamers (α, β, γ) and decipher their mechanisms of propagation and interaction with MM and MP targets. Π: residual surface discharge, Δ: dark domains where leaders are assumed to be formed, Θ: primary ionizing event in the inter-target region, Φ: filamentary discharge.*

## Duration between two Consecutive Streamers

Since a high-voltage pulse can only generate a single positive guided streamer on its rising edge, the duration between two consecutive streamers (positive and guided) is $T_G = 1/f_{rep} = 200$ μs in a repeatability situation. However, it is expected to be slightly higher in a self-organizing situation since $v_α > v_β > v_γ$. These durations between two consecutive streamers can be measured from **Figure 7** which, at first sight, indicates values that are much shorter than 200 μs. Hence, the duration between the α and β streamers on the MM target is estimated to be $1447 − 1358 = 89$ ns while the duration between the β and γ streamers is about $1550 − 1447 = 103$ ns. For each of these values, it is a mandatory to add the $T_G$ period because the ICCD camera triggers an acquisition on each rising edge, thus every 200 μs, whether for the α, β or γ streamers. The principle of this artifact is illustrated in **Figure 8** where:

- **Figure 8e** shows α streamers that could be obtained by summing only the ICCD acquisitions triggered on the −600 μs, 0 μs, 600 μs and 1200 μs rising edges of **Figure 8a**.
- **Figure 8f** shows β streamers that could be obtained by summing only the ICCD acquisitions triggered on the −400 μs, 200 μs and 800 μs rising edges of **Figure 8b**.
- **Figure 8g** shows γ streamers that could be obtained by summing only the ICCD acquisitions triggered on the −200 μs, 400 μs and 1000 μs rising edges of **Figure 8c**.

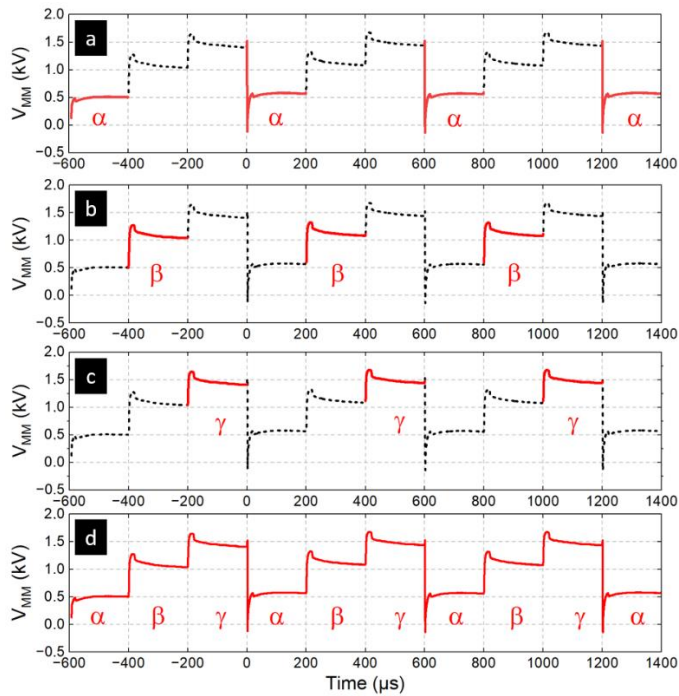

*Figure 8. Time profiles of the voltage measured in the MM target considering (a) α-phase highlighted, (b) β-phase highlighted, (c) γ-phase highlighted, (d) all phases highlighted. Schematic view of a train of (e) α-streamers, (f) β-streamers, (g) γ-streamers. (h) Fast ICCD imaging picture of three self-organized guided streamers trains.*

Considering all the rising edges of the oscillogram reported in **Figure 8d**, the ICCD imaging software overlays the previous acquisitions to form a single picture, as shown in **Figure 8h**. The exact durations separating consecutive streamers are reported in **Table 3**, distinguishing between repeatability physics and self-organization physics.

| Consecutive streamers | Repeated Guided Streamers | | Self-Organized Guided Streamers | |
|---|---|---|---|---|
| | Absolute Times | Relative Times | Absolute Times | Relative Times |
| α-β | 200 000 ns | 0 ns | 200 089 ns | 89 ns |
| β-γ | 200 000 ns | 0 ns | 200 103 ns | 103 ns |

*Table 3. Duration between consecutive streamers when impinging the MM target, considering absolute/relative times, repeatability/self-organization physics. Values are provided with an accuracy of ±1 ns (ICCD camera).*

### 3.2.2. Transient Emissive Phenomena Induced by the Self-Organized Positive Guided Streamers

Dark Domains

A positive guided streamer that impinges the MM target leaves a positive charge ($q_{Str}$) on it. Then, this charge can induce two types of electrical phenomena in its vicinity, especially in the inter-target region: an emissive phenomenon (electrical discharge of cold plasma) or a non-emissive phenomenon (space charge region or leader). In **Figure 7**, when the α streamer reaches the MM target (x = 110 mm, t = 1360 ns), no optical emission is instantaneously detected in the inter-target region, hence defining a dark region Δα comprised between the arrivals of the α and β streamers on the MM target, i.e., in the region where x = 110–113 mm and t = 1360–1450 ns. Similarly, when the β streamer reaches the MM target, it does not give rise to any emissive phenomenon in the inter-target region, hence the dark domain Δβ. The same observation applies to the γ streamer when it reaches the MM target: while the instantaneous ignition of a plasma discharge could have been expected in the inter-target region, it only appears at 1620 ns, i.e., nearly 100 ns later. The Δγ dark domain is therefore defined for x = 110–113 mm and t = 1550–1620 ns, as shown in **Figure 7**. These dark domains should not be interpreted as evidence of electrical inactivity; on the contrary, they can illustrate the existence of non-emissive phenomena such as space charge regions or leaders, capable of subsequently triggering ionizing phenomena. The existence of these non-emissive phenomena is demonstrated by electrical conductivity measurements of the air trapped in the inter-target region and evaluated in Section 3.3.4.

Residual Surface Discharge (Outlet Capillary)

While the head of the γ streamer reaches the MM target, its tail remains attached at the capillary exit, as photographed at 1540 ns, in **Figure 6**. This residual surface discharge remains stationary and then fades away in approximately 100 ns. It is also represented in the space–time emission diagram (see by its domain of existence Π in **Figure 7**), hence exhibiting a longer lifetime (1620 ns) than the head of the γ streamer itself (1550 ns). The region comprised between Π and Δγ (100–100 mm, 1550–1620 ns) is not considered as a dark region of interest since no emissive phenomena are detected afterwards.

Primary Ionizing Event and Filamentary Discharge in the Inter-Target Region

The first ionizing event in the inter-target region, Θ, is photographed at 1620 ns in **Figures 6** and **7**. It appears between the two targets without contacting either of them. It is followed by similar events, punctual and unstable, until 1630 ns. It is only from 1640 ns that a cold plasma discharge (Φ) propagates from the MM target to reach the MP target at 1660 ns. Two hypotheses can be considered as to the nature of this filamentary discharge:

- Φ could be a streamer, according to the definition recalled in Section 1.1. Nevertheless, the hypothesis of a guided streamer is unlikely because of the absence of a strongly emissive head and because of its slow and stationary disappearance (between 1760 and 1900 ns). Alternatively, the hypothesis of an unguided streamer seems also unlikely because of its linear and stable trajectory. Recall that the ICCD camera makes it possible to observe not a single streamer but a train of 25 000 streamers overlaid on each other in the same photograph. If Φ corresponded to a train of unguided streamers, then it could not have such a filamentary aspect as obtained in **Figure 6**. Finally, the inter-target region is characterized by a p.d product of value 250 Torr.cm, which induces an electric field too weak to generate a streamer. As an alternative, the filamentary discharge may be assimilated to an evanescent glow discharge.
- Φ could be a return stroke [36]. If so, this implies the simultaneous formation of an upward positive leader and a downward negative leader. The upward positive leader would result from an accumulation of positive charges at a very localized location of the MM target (e.g., on the few metal meshes impacted by the α, β and γ streamers). This leader would hardly propagate in the inter-target region but would remain localized in the interstitial spaces of the MM target. Simultaneously, the downward negative leader would result from the negative charges provided by the MP target which is grounded via the 1.5 kΩ resistor. Indeed, the ground has an electrical potential always maintained at 0 V and behaves as a reservoir containing an infinite number of free electrons. These electrons can therefore provide negative charges to form a downward negative leader propagating from MP to the MM target, until it reaches the upward positive leader. Then, the junction point between these two reversed polarized channels would lead to the formation of a return stroke, characterized by a uniform emission in the course of time, either during its propagation from MM to MP or during its progressive disappearance between the two targets.

The propagation velocity of the filamentary discharge is estimated at $v_{pr} = \frac{3\ mm}{1660\ ns - 1630\ ns} = 100\ km/s$. From 1660 ns to 1860 ns, i.e., during at least 200 ns, the two targets are thus connected by the discharge, while its optical emission decreases simultaneously.

## Optical Emission Spectrum in the Plasma Jet

The light of the plasma jet is characterized by optical emission spectroscopy, which shows differences in both the nature and intensity of the detected species, depending on whether the optical fiber is placed before (**Figure 9a**) or after (**Figure 9b**) the MM target. Unsurprisingly, the region before the MM target exhibits lines and rovibrational bands of $N_2$, $N_2^+$, OH, $H_\alpha$, He and O (triplet), especially because of the interaction between the jet of cold plasma and the ambient air ($H_2$, $O_2$, water vapor). In contrast, the spectrum of the inter-target region is dominated solely by bands of the second positive system of molecular nitrogen. No line of helium is measured while this vector gas supplies the CHEREL at 1 slm. The chemical composition of the filamentary discharge is thus very different from that of the α, β and γ streamers.

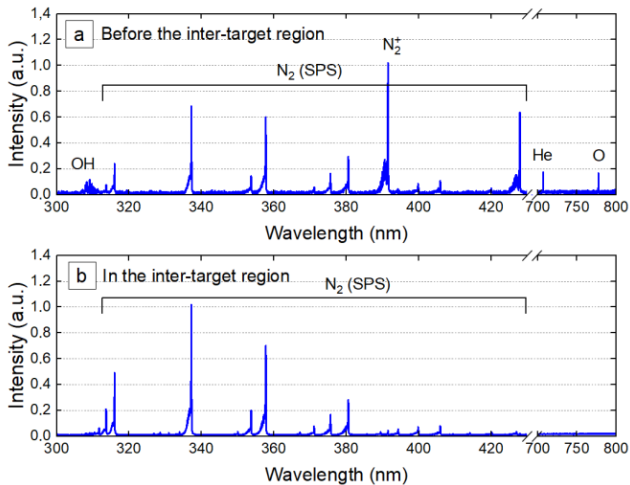

**Figure 9.** Optical emission spectra of the plasma jet measured (a) before the inter-target region at x = 105 mm and (b) in the inter-target region, at x = 112 mm.

### 3.2.3. Negative Guided Streamers

Negative guided streamers are observed by fast ICCD imaging, each streamer being generated on the falling edge of a voltage pulse. **Figure 10** shows their propagation at different times, from 0 ns (ignition) to 1320 ns (disappearance). Each negative guided streamer has a head that is more emissive than its tail, with the tail always remaining connected to the high-voltage electrode as the streamer propagates toward the MM target. Negative guided streamers manage to pass through the current probe at t = 600 ns, even though the probe strongly attenuates their optical emission. Whether before or after the current probe, the negative guided streamers form a single train: they follow each other, without inducing more complex physical phenomena like self-organization. At 1000 ns, the negative guided streamers are so weak that they can no longer be detected: they have only traveled about 7 cm of the capillary whose total length is 10 cm. Since they disappear in the capillary without ever being able to leave it, no more emission is detected at 1320 ns.

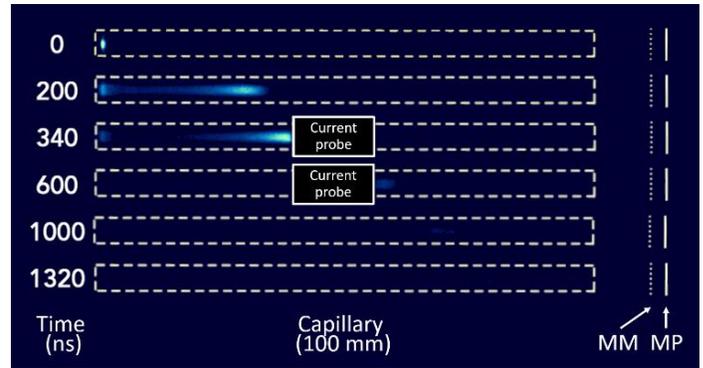

**Figure 10.** Fast ICCD pictures of negative guided streamers propagating in the CHEREL's capillary (A = 5 kV, $f_{rep}$ = 5 kHz, $D_{Cycle}$ = 10%).

## 3.3. Electrical Characterization (Ω = 3)

### 3.3.1. Overview

The high-voltage generator delivers pulses with an amplitude of 5 kV, a duty cycle of 10% and a repetition frequency of 5 kHz: hence, a repetition period of 200 μs and a resulting pulse width of 10% × 200 μs = 20 μs. The time profile of this voltage ($V_G$) is similar to a square wave signal with alternating low and high states over a range of 2000 μs, as shown in **Figure 11a**. Simultaneously, the voltage of the MM target ($V_{MM}$) is measured versus time and plotted in **Figure 11b**. This voltage shows three main discrepancies compared with $V_G(t)$, as highlighted in **Table 4**.

| | $V_G(t)$ | $V_{MM}(t)$ |
|---|---|---|
| **Number of non-zero states** | 1 | 3 states labeled α, β and γ at 0.5 kV, 1.1 kV and 1.4 kV, respectively |
| **Period** | $T_G$ = 200 μs | $T_\Omega$ = 3. $T_G$ = 600 μs (self-organization period) |
| **Profile** | Square-like signal | Succession of linear and non-linear variations |

**Table 4.** Main differences between generator voltage and voltage measured on the MM target.

From **Figure 11c**, the MP target voltage shows alternating positive and negative peaks that appear on the rising and falling edges of $V_G(t)$, respectively. Over the same period $T_\Omega$, six peaks are generated, three of which are positive and three negative. The three positive peaks, separated by $T_G$ from each other, obey the following rule: the first peak has a value of 400 V, while the next two have values of 60 V. The peak at 400 V is to be compared with $V_{MM}$ = 1500 − (−150) = 1650 V: the highest voltage drop measured on the MM target in **Figure 11b**. The guided streamers generated on the rising edges (corresponding to the α+, β+ and γ+ states) can self-organize according to a period $T_\Omega = 3.T_g$ (corresponding to the sum of the phases α, β et γ).

The three negative peaks, separated by $T_G$ from each other, have the same amplitude of only −25 V. Since the negative guided streamers are ruled by repeatability and no more by self-organization schemes, the notation for α−, β− and γ− states will not be used in the rest of the article.

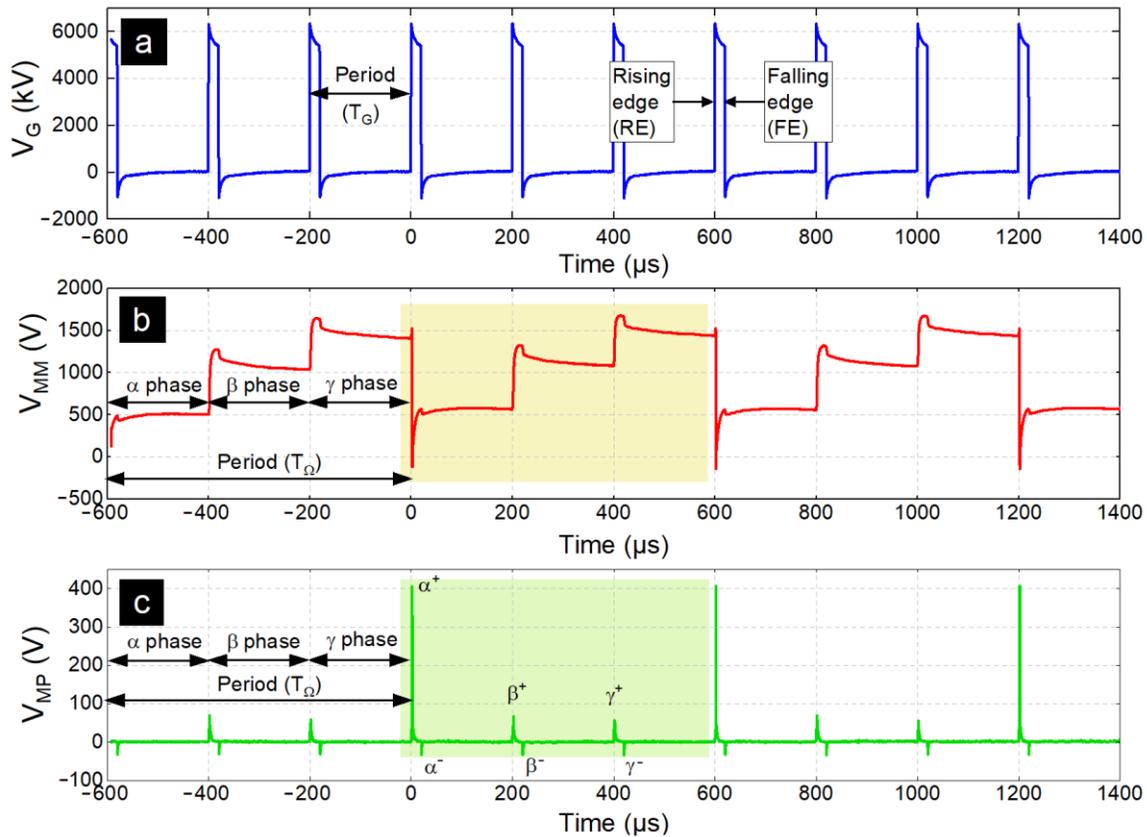

*Figure 11. Time profiles of (a) the voltage delivered by the generator with the following set parameters: A = 5 kV, f_rep = 5 kHz, D_Cycle = 10 %, (b) the voltage measured on the metal mesh target, (c) the voltage measured on metal plate target.*

### 3.3.2. Analysis of the Rising Edges

**Voltages**

The MM target voltage is plotted in **Figure 12a** over a period $T_\Omega = 600$ μs, distinguishing the α (black), β (red) and γ (green) phases initiated at $t_\alpha = 200$ μs, $t_\beta = 400$ μs and $t_\gamma = 600$ μs, respectively, i.e., initiated on the rising edges of $V_G(t)$: RE$_\alpha$, RE$_\beta$ and RE$_\gamma$, respectively. The voltage of the MM target increases over two steps and drops sharply at the end of the γ phase. These important variations are highlighted in **Figure 12b**, which plots $V_{MM}(t)$ on a relative time scale from −1 μs to +6 μs. In this figure, the rising edge of each phase and thus the instants $t_\alpha$, $t_\beta$ and $t_\gamma$ are brought to 0, so that the same reference instant ($t_{ref}$) is considered. The sharp drop in voltage occurring at the very beginning of the α phase has a value of 1530 − 130 = 1660 V obtained in only 700 ns. Then, whatever the phase, the $V_{MM}$ amplitude increases following a non-linear profile that corresponds to a capacitive loading of the MM target.

As a reminder, the MM and MP targets can be considered as the electrodes of a same capacitor (modeled by a series RC circuit) whose dielectric medium is the ambient air, as represented in **Figure 1**. The variations of the potential of the MM target can thus change the voltage measured in the MP target, as shown in **Figure 12c**. Unsurprisingly, the α phase is marked by the stronger $V_{MP}$ peak with a value as high as 40 V centered at 1.9 μs. This peak is directly related to the sharp drop of $V_{MM}$ (1660 V) over the interval 1.5–2.2 μs (**Figure 12b**). Subtracting $V_{MP}(t)$ to $V_{MM}(t)$ as given by relation (1) provides the voltage in the inter-target region ($V_{inter}$), as plotted in **Figure 12d**. It turns out that this voltage shows a temporal profile very close to that of $V_{MM}(t)$ because the values of $V_{MP}(t)$ are comparatively very small, of the order of a few tens of Volt.

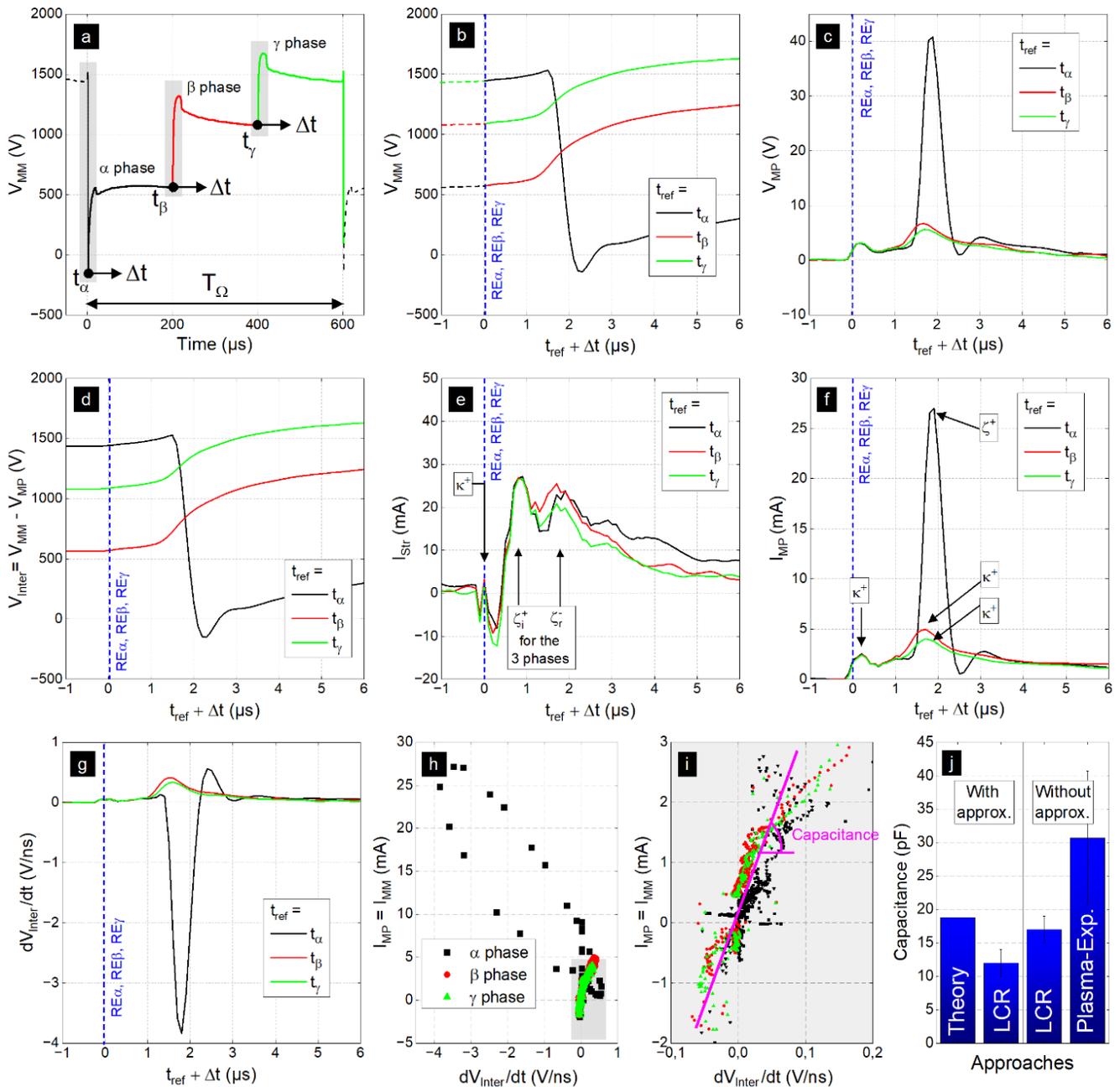

*Figure 12. (a) Time profile of the voltage measured in the MM target ($V_{MM}$) upon its interaction with $\Omega = 3$ self-organized guided streamers. (b–g) Time profiles of electrical signals sequenced in α, ß and γ phases starting at $t_\alpha = 0$ μs, $t_\beta = 200$ μs and $t_\gamma = 400$ μs. Each of these instants corresponds to the rising edge (RE) of a phase along a $T_G$ period. (b) Voltage in the MM target (c) voltage in the MP target, (d) Voltage in the inter-target region, (e) current in the capillary, (f) current in the MP target, (g) derivative of the voltage in the inter-target region. (h) Electrical current measured in the experimental setup vs. the derivative of $V_{inter}$. (i) Magnification view of (h) to estimate the capacitance in the inter-target region. (j) Capacitance in the inter-target region assessed following different approaches (theory, LCRmetry and exp. deduction from (j)) with approximation (metal mesh replaced by a metal plate) and without.*

## Currents

A current probe is placed to assess the streamer current $I_{Str}(t)$ in the CHEREL capillary while another is used to assess the current $I_{MP}(t)$ collected by the MP target. The time profile of $I_{Str}(t)$, plotted in **Figure 12e**, is identical whatever the phase (α, β or γ) and always consists of three peaks:

- A capacitive current peak, noted κ, appearing at the rising edge and proportional to the derivative of the supplied pulse voltage. The exponent " + " means that the peak is positive.
- A first conduction current peak, noted $\zeta_i^+$. This peak corresponds to the incident positive guided streamer.
- A second conduction current peak, noted $\zeta_r^-$. This peak corresponds to the negative guided streamer, reflected by

the MM target and which therefore arrives with a delay of 1 μs compared to the first. The mechanisms governing this counter-propagation phenomenon are detailed in our previous article [15].

Since the ICCD pictures of **Figure 6** show that the three guided streamers follow one another, the three time-shifted $\zeta_i^+$ peaks would be expected in **Figure 12e**. The fact that they all appear at $t_{ref}$ + 0.9 μs is that the current probe is located at x = 50 mm: a location where self-organization has not yet spontaneously emerged (see **Figures 1** and **6**). This explains why the current profile in the capillary is different from the current measured in the MP target, as shown in **Figure 12f**. The main current peak of $I_{MP}(t)$ reaches a value close to 27 mA for the α phase against only 5 mA for the two others. Considering these figures, several remarks are noteworthy:

- The conservation of the electric charge is respected at all points of the experimental setup, from the CHEREL to MM and from MM to MP. In the CHEREL, the profile of the current is always the same from one phase to another, so that the conservation hypothesis is verified taking $T_G$ (200 μs) as the reference period. However, in the inter-target region, the current shows a different temporal profile so that the relations (2) and (3) are respected. Therefore, the conservation of the electric charge is validated only upon $T_O$ (600 μs) considered here as the reference period.
- It is important to rule on the capacitive or conductive nature of the current peaks obtained in **Figure 12f** because only a conductive current peak can be associated with a streamer. Conversely, a capacitive current peak has the characteristic of being proportional to the derivative of the voltage, the proportionality coefficient corresponding to the capacitance of the dielectric medium as indicated in relation (4). By plotting the derivative of $V_{Inter}(t)$ on the α, β and γ phases, **Figure 12g** discriminates on the one hand the capacitive current peaks, which are positive and which correspond to those of the β and γ phases, and on the other hand the conduction current peak, which is oriented in the opposite direction and whose amplitude is in absolute value 10 times greater. Conduction and capacitive current peaks can also be differentiated by plotting $I_{MP}(t)$ vs. $V_{Inter}/dt$, as proposed in **Figure 12h**. According to (4), this figure represents a capacitive current as a straight line whose slope provides the capacitance while a conductive current is represented by a closed contour whose area provides its energy. Here, the closed contour (black curve) indicates the existence of a conduction current peak upon the α phase while the linear profiles of the β and γ phases indicate capacitive current peaks. For the sake of clarity, a magnification view of **Figure 12h** is proposed in **Figure 12i** for $dV_{Inter}/dt$ between −0.1 and 0.2 V/ns. A graphical reading provides a capacitance of 31 (±9) pF.
- In **Figure 12f**, at $t_{ref}$ + 0.3 μs, each phase shows a first peak of 2–3 mA, which corresponds to a capacitive peak, denoted $\kappa^+$. At $t_{ref}$ + 1.7 μs, each phase is characterized by a second current peak: (i) the peaks of the β and γ phases are relatively small (4-5 mA) and correspond to capacitive peaks, also referenced $\kappa^+$; (ii) The peak of the α phase has a magnitude as high as 27 mA and corresponds to a flow of free charges that propagates into the inter-target region, forming the filamentary discharge already detected by fast ICCD imaging (see **Figure 6**). The α peak is therefore a conduction peak noted $\zeta^+$. Finally, a third peak seems to be detected at $t_{ref}$ + 2.9 μs, although there is no evidence to suggest whether this is a simple artifact or a capacitive peak. No third peak is detected for the β and γ phases.

- Since the MP target is connected to ground via a 1.5 kΩ resistor, the value measured by the current probe in **Figure 12f** can be checked by applying Ohm's law. It turns out that multiplying 27 mA by 1.5 kΩ provides 40.5 V: a value that corresponds with the one measured in **Figure 12c**.

The capacitance value of 31 (±9) pF is compared to those obtained by three other approaches, as shown in **Figure 12j**. The theoretical approach is based on Equation (5), which assumes the parallel connection of two flat and solid electrodes. A value of 19 pF is then obtained with, however, the bias that one of the two electrodes is not full but a mesh (i.e., containing interstitial spaces). We also used an RLC meter to measure the capacitances of the MM/air/MP and MP/air/MP systems (with the same characteristics as mentioned in Section 2.1). Overall, the three comparative approaches lead to a result similar to the one deduced in **Figure 12i**.

$$V_{Inter}(t) = V_{MM}(t) - V_{MP}(t) \quad (1)$$

$$I_{Str}(t) \neq I_{MP}(t) \quad (2)$$

$$I_{Inter}(t) = I_{MM} = I_{MP}(t) \quad (3)$$

$$I_{Inter}(t) = C. \frac{dV_{Inter}(t)}{dt} \quad (4)$$

$$C = \varepsilon_0.\varepsilon_r.\frac{S}{d} \quad (5)$$

$$E_{ext}(\Omega) = \frac{V_G - V_{MM}(\Omega)}{110 \; mm} \quad (6)$$

Electric Fields

The spontaneous emergence of self-organized guided streamers requires that the CHEREL interacts with a target issued to a floating potential: the MM target (**Figure 4**). During each phase, a streamer brings its electric charge $q_{Str}$ on the MM target whose potential is initially −200 V, 580 V and 1100 V for the α, β and γ phases, respectively (**Figure 12a**). As the voltage applied on each rising edge is always $V_G$ = 5 kV, this means that the external electric field applied to the α streamers is different from that applied to the β streamers, which is itself different from that applied to the γ streamers. According to Equation (6), the α, β and γ streamers are thus subjected to external electric fields of 473 V·cm⁻¹, 402 V·cm⁻¹ and 354 V·cm⁻¹, respectively. These decreasing electric field values explain why $v_α > v_β > v_γ$, as already reported in **Figure 7** and in **Table 2**. The β streamers propagate in the ionic trace of the α streamers, i.e., in a weakly ionized channel where the external electric field is less intense, hence $v_β < v_α$. The same reasoning applies for the γ streamers, which also propagate in an even less intense electric field than the previous one, hence $v_γ < v_β$. As already mentioned in Section 3.2.1, the three self-organized guided streamers gain in velocity at the capillary's outlet while keeping the same inequality ratios ($v_\alpha > v_\beta > v_\gamma$). These higher velocities may result from a gaseous environment different from that of the capillary, producing non-negligible densities of $N_2^+$ ions and electrons by Penning ionization, as detailed in **Figure 9**. These charged particles are indeed more sensitive to the electric field and, therefore, could sustain faster streamer propagation.

### 3.3.3. Analysis of the Falling Edges

**Figure 13a** shows that during the α, β and γ phases, $V_{MM}(t)$ can change as a function of time but in relatively close proportions. Thus, $V_{MM}(t)$ does not present a sharp drop during the α phase, as was the case on REα in **Figure 12f**. This means that the negative guided streamers generated upstream of the MM target cannot give rise to the filamentary discharge in the inter-target region. This also explains why the voltage of the MP target remains almost unchanged as shown in **Figure 13b**, where voltage values never exceed −4 V. Therefore, one can reasonably consider that the voltage across the MM/Air/MP region is such that $V_{inter}(t) \approx V_{MM}(t)$. While the voltage variations remain relatively small, those of the current through the capillary are much larger. As shown in **Figure 13c**, the time profile of $I_{Str}$ does not depend on the phase and always has a negative capacitive current peak ($\kappa^-$) detected on the falling edge (FE), followed 1 μs later by a negative conduction current peak ($\zeta^-$). Unlike the previous case (**Figure 12e**), no second conduction current peak is detected. This means that the negative guided streamers can propagate for a distance equal to at least 50 mm (distance between the current probe and the HV electrode) but do not have enough energy to reach the MM target, hence the absence of counter-propagation. This result, which is based on electrical measurements, is confirmed by the characterization performed by fast ICCD imaging in Section 3.2.3 (**Figure 10**).

Even if cold plasma is not present in the inter-target region and over such a short period of time, the MP target can be subjected to different electrical phenomena. Indeed, $V_{MP}(t)$ varies between 0 and −4 V (**Figure 13b**) while $I_{MP}(t)$ varies between 0 and −2 mA (**Figure 13d**). The first peak, closest to the falling edge, corresponds to the capacitive current peak ($\kappa^-$). At first sight, the nature of the second current peak is more speculative owing to the detection limits of the electrical probes. By deriving $V_{inter}$ with respect to time as in Section 3.3.2., **Figure 13e** reveals that the peaks of the three phases are all oriented downward. However, in the case of **Figure 12g** (rising edges), the conduction current peak is oriented in the opposite direction to the capacitive currents. Moreover, **Figure 13f**, which plots $I_{MP}$ as a function of $dV_{inter}/dt$, shows that there is no clear closed contour. Finally, it is worth reminding that fast ICCD imaging did not reveal any guided streamer counter-propagating from the MP target. All these reasons support that the second current peak in **Figure 13d** is a capacitive current peak.

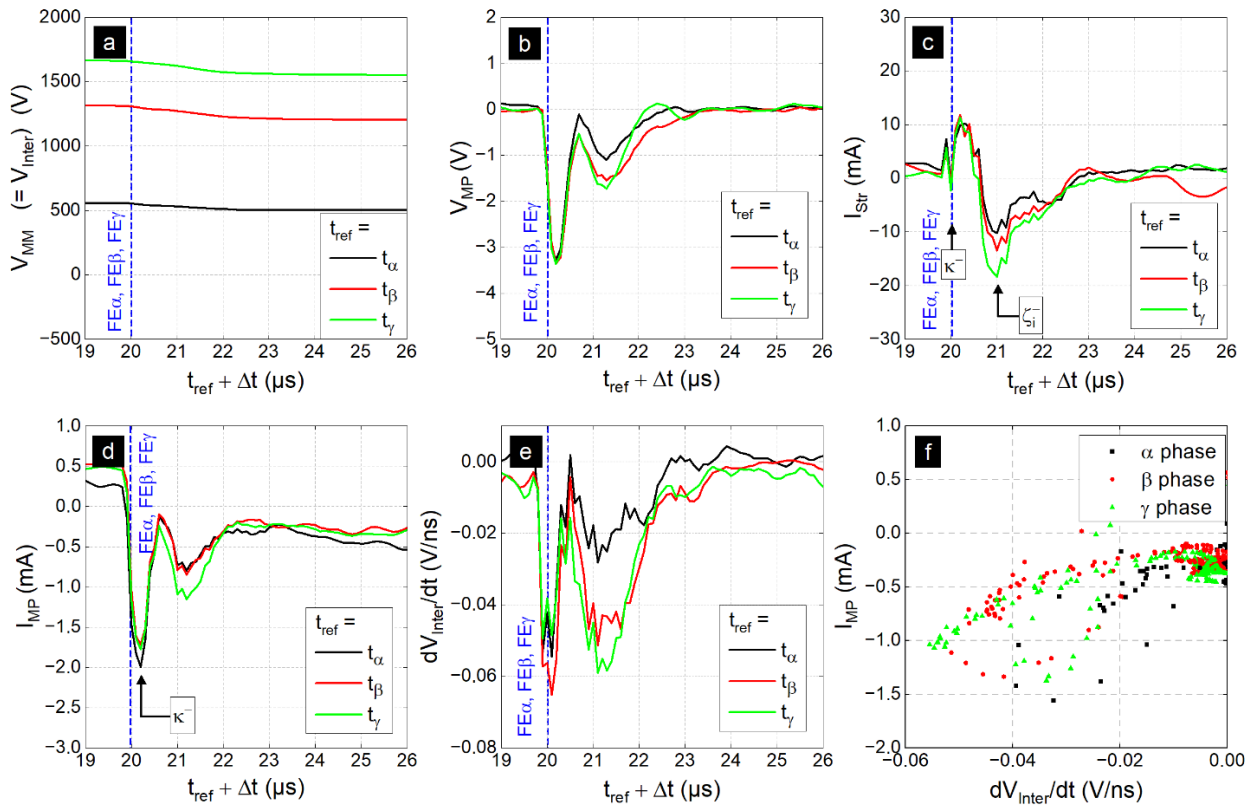

*Figure 13. Time profile of the voltage and current associated to the self-organized guided streamers with Ω = 3. Each profile is sequenced in α, β and γ phases starting at $t_\alpha = 0$ μs, $t_\beta = 200$ μs and $t_\gamma = 400$ μs, respectively. Each of these instants corresponds to the rising edge (RE) of a phase along a $T_\Omega$ period. (a) Voltage in the MM target, (b) voltage in the MP target, (c) current in the capillary, (d) current in the MP target, (e) derivative of $V_{inter}$, (f) electrical current measured in the experimental setup vs. the derivative of $V_{inter}$.*

### 3.3.4. Understanding How Streamer Self-Organization Triggers Electrical Phenomena in the Inter-Target Region

#### Electrical Charges Carried by the Streamers and Deposited on the MM Target

As we demonstrated in Section 3.1 (**Figure 4**), the condition that allows guided streamers to self-organize is the presence of an MM target carried at a floating potential. Whether α, β or γ, a streamer carries an electric charge ($q_{Str}$) whose value can be determined by measuring the area under its current peak ($I_{Str}$), as proposed by Equation (7). As an example, during the beta phase, the current peaks on the rising and falling edges are shown in **Figure 14a,b** respectively. The integration of the area of these peaks allows us to determine the electric charge $q_{Str}^+$ (carried by the positive guided streamer of the rising edge) and the electric charge $q_{Str}^-$ (carried by the negative guided streamer of the falling edge).

Generalizing these measurements to the three phases of a single 600 µs $T_\Omega$ period, **Figure 14c** presents the values of the $q_{Str}^+$ and $q_{Str}^-$ charges generated in the CHEREL capillary. The algebraic sum of these charges, denoted QΩ, is given by Equation (8). Thus, for the "rising edge" condition, **Figure 14c** shows electrical charges of about 24, 15, and then 12 nC for the streamers associated with the α, β and γ phases, respectively. Conversely, the "falling edge" condition is characterized by negative streamers with individual electrical charges of no more than 3 nC. Over a $T_\Omega$ period of 600 µs, the sum of the positive charges is 51.7 ± 1.0 nC while the sum of the negative charges is −4.8 ± 0.1 nC, leading to an overall net charge $Q_{\Omega,\text{Cherel}} = 46.9\ nC$ at the x = 50 mm position (**Figure 1**). This net charge does not correspond to the charge accumulating on the MM target since the fast ICCD imaging characterizations showed that the negative guided streamers did not exit the CHEREL (**Figure 10**). Therefore, the overall net charge received by the MM target at the end of the $T_\Omega$ period is: $Q_{\Omega,\text{MM}} = q_{Str}^{\alpha+} + q_{Str}^{\beta+} + q_{Str}^{\gamma+} = 51.7\ nC$.

#### Characteristics of the Inter-Target Region

The two targets are separated by air (dielectric medium), thus forming a series RC circuit, characterized by four parameters:
- The constant τ, given by the relation (9),
- The electrical conductivity $\sigma_{Inter}$ of the air, given by the relation (10),
- The maximum electric charge $Q_{Inter}^{max}$ of the inter-target region, given by Equation (11)
- The voltage $V_{Inter}$ whose temporal profile is shown in **Figure 15**.

Before analyzing this figure in detail, it should first be noted that $V_{Inter}(t)$ presents not one but three capacitive rises (or three phases) before undergoing an abrupt drop corresponding to the breakdown of the gas in the inter-target region. By graphical reading, the breakdown potential from which the filamentary discharge Φ can propagate is $V_{Br} = V_M − V_A = 1530\ V − (−150\ V) = 1680\ V$. Since $C_{Inter} = 28\ pF$ (result obtained from **Figure 12**), Equation (11) allows us to deduce the value of $Q_{Inter}^{max}$ and to evaluate it at 47 nC.

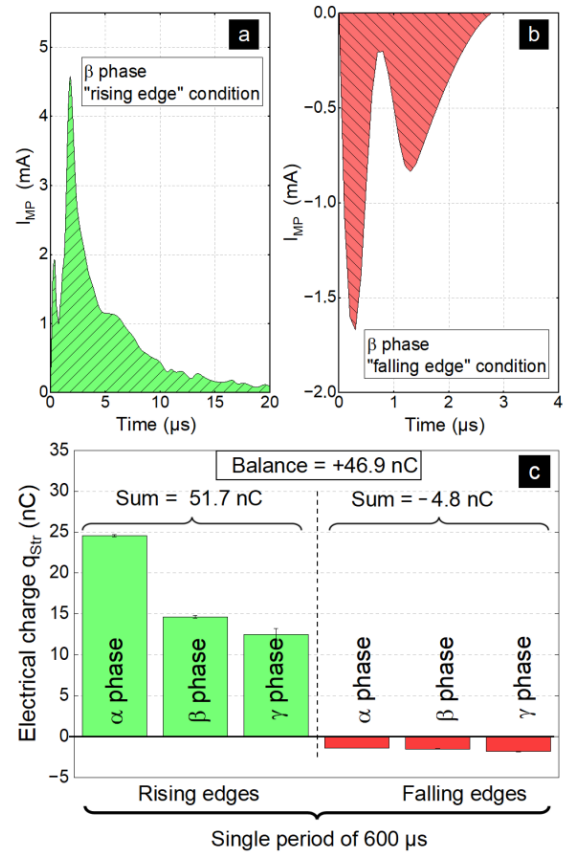

**Figure 14.** Time profile of the current associated to a β-streamer generated (a) on the rising edge of the high-voltage pulse and (b) on the falling edge of the high-voltage pulse. (a,b) Each hatched area corresponds to the electrical charge associated with the current peak. (c) Electrical charges of the six guided streamers generated upon a same $T_\Omega$ period, considering rising/falling edges as well as α, β and γ phases. In all cases: A = 5 kV, $f_{rep}$ = 5 kHz, $D_{Cycle}$ = 10%.

#### Maximum Capacitive Load Condition—Gas Breakdown Condition

From the results in **Figure 14**, the electrical charge of a single streamer, $q_{Str}$, is always lower than $Q_{Inter}^{max}$. In other words, as long as the $q_{str}$ charges accumulating on the MM target have a value $Q_{\Omega,MM}$ lower than $Q_{Inter}^{max}$, Equation (12) is not verified. Therefore, no filamentary discharge Φ can be generated in the inter-target region and thus $V_{Inter}(t)$ follows the capacitive loading law given by Equation (13). On the other hand, when $Q_{\Omega,MM} = 51.7\ nC$ (total charge of the accumulated streamers on the MM target) exceeds the value $Q_{Inter}^{max} = 47nC$, then Equation (12) is satisfied and the filamentary discharge Φ can propagates through the inter-target region. This is demonstrated by fast ICCD imaging in **Figure 6** and by current probe in **Figure 12f** (conduction current peak). The emergence of the filamentary discharge Φ is conditioned by an electric charge inequality (Equation (12)) which can also be expressed as a voltage inequality. Indeed, $Q_{\Omega,MM}$ increases as the α, β and γ streamers reach the MM target, which induces an increase in $V_{Inter}(t)$ as indicated by Equation (14) and confirmed by **Figure 15**. It is only if Equation (15) is satisfied, i.e., $V_{Inter}$ becomes higher than the breakdown potential $V_{Br}$, that the filamentary discharge Φ appears.

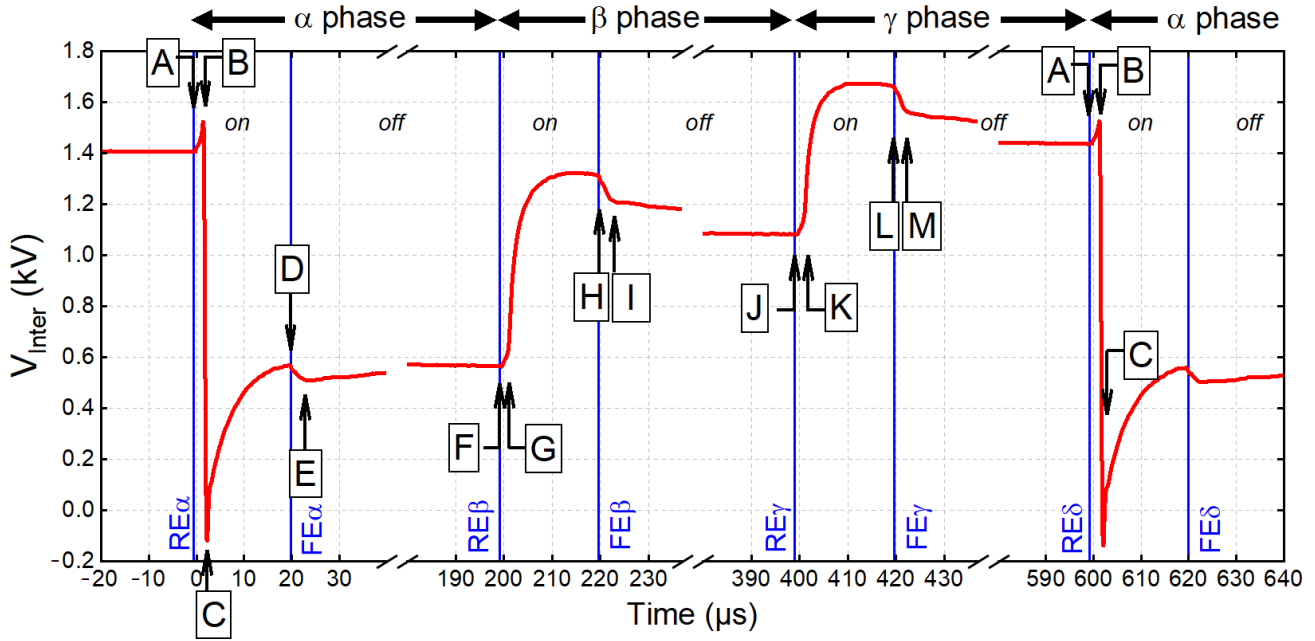

*Figure 15. Time profile of the voltage measured in the inter-target region. The three phases correspond to the self-organization of three guided streamers (A = 5 kV, $f_{rep}$ = 5 kHz, $D_{Cycle}$ = 10 %). RE and FE stand for rising edge and falling edge, respectively.*

$$q_{Str} = \int_{\tau_{Str}} I_{Str}(t) . dt \tag{7}$$

$$Q_\Omega = \underbrace{\sum_{k=1}^{N} q_{Str}^+}_{rising\ edges} + \underbrace{\sum_{k=1}^{N} q_{Str}^-}_{falling\ edges} \tag{8}$$

$$\tau = R_{Inter} . C_{Inter} \tag{9}$$

$$\sigma_{Inter} = \frac{L}{R_{Inter} . S} = \frac{3 \times 10^{-3}_{(m)}}{R_{Inter} \times 64 \times 10^{-3}_{(m^2)}} = \frac{0.0469}{R_{Inter}} \tag{10}$$

$$Q_{Inter}^{max} = C_{Inter} . V_{br} \tag{11}$$

$$Q_\Omega \geq Q_{Inter}^{max} \tag{12}$$

$$V_{Inter}(t) = V_{max} . \left[ 1 - e^{-\frac{t}{\tau}} \right] \tag{13}$$

$$V_{Inter}(t) = \frac{Q_\Omega(t)}{C_{Inter}} \tag{14}$$

$$V_{Inter}(t) > V_{br} \tag{15}$$

Analysis of $V_{Inter}(t)$

The time profile of $V_{Inter}(t)$ in **Figure 15** shows that each phase starts with a voltage rise of some 100 V over a very short time, $t_{delay}$, of the order of 2–3 µs (portions AB, FG and JK). $t_{delay}$ corresponds to the lifetime of a positive guided streamer, as highlighted in **Figure 6**.

Immediately after and only during the α phase, a sharp linear drop of $V_{Inter}$ is observed from point B to point C: −0.13 − (1.53) = − 1.66 kV in 700 ns, as already mentioned.

Then, whatever the phase, $V_{Inter}$ presents a capacitive loading profile as established by Equation (15): an initially very fast rise that slows down and converge asymptotically to a limit value (CD, GH and KL portions). These capacitive loads are shifted by $t_{delay}$ with respect to the rising edges (RE). By fitting the parameters of this equation to the $V_{Inter}(t)$ curve, it is possible to derive the time constants τRC, as carried over in **Table 5**. Since the capacitance of the MM/air/MP region depends only on geometrical parameters, it can be considered constant and of value equal to 28 pF from **Figure 12j**. Moreover, relations (12) and (13) provide the values of resistance ($R_{Inter}$) and conductivity ($\sigma_{Inter}$) of the air respectively: during the α phase, the resistance of the air is thus estimated to be 156 kΩ before being reduced by a factor of two whether for β or γ phase (Table 5). This result supports the hypothesis of leaders evoked in **Figure 7**, capable of forming in the dark domains (Δα, Δβ, Δγ) before the emergence of Θ: primary ionizing event in the inter-target region, which itself precedes the filamentary discharge Φ.

At the end of the 20 µs pulse, $V_{MM}(t)$ drops by some 100 V over a period of a few µs (between D and E, between H and I, between L and M) before stabilizing. Since the negative guided streamers remain confined in the capillary, they cannot be responsible for this voltage drop and therefore the time lapse in question is different from $t_{delay}$. This voltage drop of some 100 V corresponds to the beginning of a capacitive discharge whose duration will in fact extend until the next rising edge, i.e., points F, J and A for the α, β and γ phases, respectively. The filamentary discharge Φ naturally operates on much shorter time scales, following rather an ohmic decay profile (through $R_{Inter}$) than a capacitive discharge profile (through $C_{Inter}$).

The time profile of the capacitive unloading is increasing on the α phase (between E and F) but decreasing on the β (between I and J) and γ (between M and A) phases. This type of discrepancies can also be found on classical RC series circuits in analog electronics depending on whether the time constant $\tau RC$ is very short or of the same order of magnitude as $\tau unload$: the duration during which the capacitive unloading can take place is estimated here as $\tau_{unload} = (1 - D_{Cycle}) \times T_G = 0.9 \times 200\ ms = 180\ ms$ (see complementary results in Appendix A).

| Portions of the Curve in Figure 15 | $\tau_{RC}$ ($\mu s$) | $C_{Inter}$ (pF) | $R_{Inter}$ (k$\Omega$) | $\sigma_{Inter}$ ($\Omega^{-1}.m^{-1}$) |
|---|---|---|---|---|
| CD (ou DE) | 4.7 μs | 30 | 156 | $3.0 \times 10^{-6}$ |
| GH (ou HI) | 1.9 μs | 30 | 65 | $7.2 \times 10^{-6}$ |
| KL (ou LM) | 1.7 μs | 30 | 56 | $8.4 \times 10^{-6}$ |

*Table 5. Electrical parameters of the MM/air/MP region.*

# 4. Conclusions

The self-organization of guided streamers has been demonstrated by combining two methodological approaches: (i) electrical characterization by electrical probes to observe current peaks associated with individual guided streamers and (ii) optical characterization by fast ICCD camera to observe trains of 25 000 guided streamers. These two approaches are complementary and consolidate each other by reducing biases and limitations of the devices (optical artifacts, oscilloscope sampling, jitter, sensitivity, …).

While the repeatability of the guided streamers requires only that the CHEREL is supplied with high-voltage pulses delivered at a given repetition frequency, their self-organization requires not only the use of a pulse generator but also that the pulse parameters have very specific values (Table 1) and that the CHEREL interacts with two targets placed one behind the other: the MM target brought to floating potential followed by the MP target that is grounded via a 1500 Ω resistor (**Figures 3** and **4**).

Under these conditions, we have shown that only positive guided streamers can self-organize, negative guided streamers being only ruled by the physics of repeatability (**Figure 11**) and being unable to exit the capillary. In the case where $\Omega = 3$, the emergence of self-organization operates in the CHEREL capillary (x = 50 mm) and leads to the propagation of α, β and γ streamers at 75.7 km·s$^{-1}$, 66.5 km·s$^{-1}$ and 58.2 km·s$^{-1}$, respectively (**Figure 6, Figure 7**). These differences in velocity are a direct consequence of the floating potential of the MM target: when the α streamer reaches the target, it deposits an electrical charge that increases $V_{MM}$. Then, the external electric field is reduced so that the velocity of the next streamer (β streamer) is slightly decreased. The same reasoning applies to the γ streamer, and thus explains the inequality $v_\alpha > v_\beta > v_\gamma$ (Table 2). Consistently, the charge deposited by the α streamer is greater than that deposited by the β streamer, which is itself greater than that of the γ streamer (**Figure 14**).

By its intermediate position between the CHEREL and the MP target, the MM target constitutes one of the two electrodes of the

"MM/air/MP" system, which can be modeled as a series RC circuit. The self-organized guided streamers can deposit their electrical charges on the MM target until they reach a total net value $Q_{Inter}^{max} = 47\ nC$. This latter one is reached after the capacitive loading and unloading of the α, β and γ phases (**Figure 15**). It is associated with the breakdown potential of the gas confined in the inter-target region, i.e., the air of which only the rovibrational bands of $N_2$ emit a visible radiation (**Figure 9**). The electrical phenomenon generated is then a filamentary discharge Φ, which has the particularity of not being a streamer (guided or unguided) and which does not originate from one of the target electrodes but from a Θ event within the gas confined in the inter-target region (**Figures 6** and **7**). This discharge specificity may suggest the existence of leaders in the inter-target region, consecutive to each of the self-organized streamers and located dark domains (**Figure 7**). To these two transient emissive phenomena, a third one is detected: residual surface discharge.

Although our study focused on $\Omega = 3$, we also demonstrated that the emergence of self-organization could operate on lower (2) and higher (5) values of $\Omega$ (**Figure 5**).

In outlook, it is important to underline that this experimental study can have interesting spinoffs for applied research, particularly regarding the engineering of plasma medical devices. For example, in one of our previous papers, we developed a transferred plasma catheter that allows endoscopic treatment of the biliary tree of porcine models [37]. This type of catheter, supplied by helium and powered by high-voltage pulses, has a floating electrode (long metallic wire) that endoscopically transfers cold plasma to the area to be treated (human tissue, cells, …). An obvious parallel can be made by considering that this floating electrode and this region correspond to the MM and MP electrodes, respectively. Therefore, this means that the transferred plasma catheter could be used in a modality favoring the self-organization of guided streamers, potentially inducing new therapeutic modalities. The effects of self-organized plasma patterns could be tremendous, as highlighted in the works of Chen et al. where antitumor effects are strengthened in the case of breast cancer cells exposed to stratified plasma patterns [38].

**Author Contributions:** Conceptualization, H.D. and T.D.; methodology, H.D.; validation, T.D.; formal analysis, H.D. and T.D.; investigation, H.D.; resources, T.D.; data curation, T.D.; writing—original draft preparation, H.D.; writing—review and editing, T.D.; visualization, T.D.; supervision, T.D.; project administration, T.D.; funding acquisition, T.D. All authors have read and agreed to the published version of the manuscript.
**Funding:** The research presented in this article was made possible through the support of the doctoral school ED-564 (Ecole Doctorale Physique Ile-de-France) who funded the PhD contract of H. Decauchy (Contract Decauchy-2019)). The authors gratefully acknowledge the financial support provided by the INCA (Institut National du CAncer) through the "High Risk High Gain" program (Contract RHG-MP22-039).
**Institutional Review Board Statement:** Not applicable.
**Informed Consent Statement:** Not applicable.

**Data Availability Statement:** The data presented in this study are available upon request from the corresponding author. The data are not publicly available due to their large storage volume.

**Acknowledgments:** The authors would like to thank Sorbonne Université and the Île-de-France Region for supporting fundamental research in plasma physics by co-funding the PF2ABIOMEDE platform project through Sesame 2016 (Region Ile de France) and the SU platforms program.

**Conflicts of Interest:** The authors declare no conflict of interest.

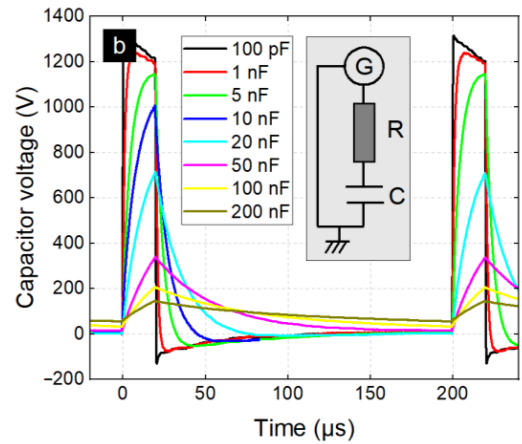

*Figure A1. (a) Capacitor voltage vs. rising time of a theoretical series RC circuit, (b) Time profile of capacitor voltage in a series RC circuit supplied with high-voltage nanopulse generator, A = 5000 V, f = 5 kHz and $D_{Cycle}$ = 10%.*

# Appendix A

**Figure A1a** represents the voltage measured across a capacitor during its loading. While this capacitive loading would theoretically take infinite time, in practice it is charged after a time equal to 5 times the rise time (trise). This time constant is defined as the time required for the voltage across the capacitor to reach a value of 63.2% of the final theoretical value.

**Figure A1b** represents the voltage measured across a capacitor placed in a series RC circuit where the resistance is fixed at 1 kΩ and the capacitance is modified to vary the time constant $\tau_{RC}$. The circuit is supplied with the high voltage pulse generator, according to the same electrical parameters used in this article: A = 5000 V, f = 5 kHz and $D_{Cycle}$ = 10%. This figure clearly shows that immediately after the falling edge, the capacitive unloading of the capacitor sees its voltage increase for C = 100 pF and 1 nF while it decreases for capacitance values greater than 10 nF. The same trends are observed in **Figure 15**: an increasing profile for the α phase and decreasing for the β and γ phases.

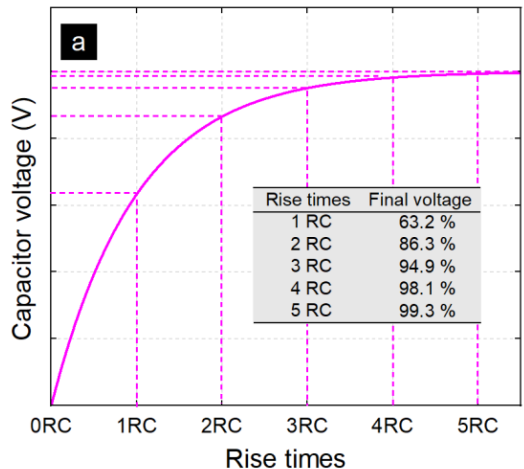

| Rise times | Final voltage |
|---|---|
| 1 RC | 63.2 % |
| 2 RC | 86.3 % |
| 3 RC | 94.9 % |
| 4 RC | 98.1 % |
| 5 RC | 99.3 % |